\documentclass[prl,aps,twocolumn]{revtex4}
\usepackage{graphicx}
\usepackage{epstopdf}
\usepackage{latexsym}
\usepackage{amsmath}
\usepackage{graphicx}
\usepackage{dcolumn}
\usepackage{bm}
\usepackage{wasysym}

\newcommand{\beq}{\begin{equation}}
\newcommand{\eeq}{\end{equation}}
\newcommand{\beqn}{\begin{eqnarray}}
\newcommand{\eeqn}{\end{eqnarray}}

\begin{document}

\title{A square lattice algebraic spin liquid with SO(5) symmetry}

\author{Cenke Xu}
\affiliation{Department of Physics, Harvard University, Cambridge
MA 02138, USA}

\author{Subir Sachdev}
\affiliation{Department of Physics, Harvard University, Cambridge
MA 02138, USA}

\date{\today}

\begin{abstract}
We propose a critical spin liquid ground state for $S=1/2$
antiferromagnets on the square lattice. In a renormalization group
analysis of the `staggered flux' algebraic spin liquid, we examine
perturbations, present in the antiferromagnet, which break its
global SU(4) symmetry to SO(5). At physical parameter values, we
find an instability towards a fixed point with SO(5) symmetry. We
discuss the possibility that this fixed point describes a
transition between the N\'eel and valence bond solid states, and
the relationship to the SO(5) non-linear sigma model of Tanaka and
Hu.
\end{abstract}

\maketitle

Critical spin liquids appear in a variety of contexts in studies
of correlated electrons in insulators and superconductors. These
are states in which spin rotation symmetry is preserved and there
is a gapless spectrum of spin excitations which do not have a
quasiparticle interpretation. The simplest, and best understood,
states \cite{csy,ruegg} are described by the Wilson-Fisher fixed
point of the Landau-Ginzburg-Wilson theory of fluctuations of the
vector antiferromagnetic (N\'eel) order parameter
$\boldsymbol{N}$. These provide a description of the quantum phase
transition of dimerized antiferromagnets between a state with
long-range N\'eel order and a spin-gapped state with $S=1$
quasiparticle excitations which are quanta of the field
$\boldsymbol{N}$. However, the anomalous dimension of the field
$\boldsymbol{N}$ is quite small at the critical fixed point,
implying that a perturbative description in terms of
$\boldsymbol{N}$ quasiparticles provides a reasonable description
of the zero temperarure spectrum.

A separate category of critical (or `algebraic') spin liquids
involve a description in terms of neutral $S=1/2$ excitations
(`spinons'). The first example of this was the `staggered flux'
spin liquid state of Affleck and Marston \cite{brad}, whose low
energy theory involved spinons, represented two-component massless
Dirac fermions $\Psi_\alpha$ ($\alpha = 1 \ldots N=4$ is a
combined SU(2) spin and valley index), coupled to an emergent U(1)
gauge field $a_\mu$. The spin fluctuations of this theory have
been better understood in subsequent work
\cite{rantwen,stableu1,motherasl}: as we will review below, for
$N$ sufficiently large, the low energy theory is a conformal field
theory (CFT) with a global SU(4) symmetry. Another spin liquid
state involves bosonic spinons represented by relativistic scalars
$z_p$ ($p=1,2$ is a SU(2) spin index) coupled to a U(1) gauge
field \cite{mv}. It has been proposed \cite{senthil} that this is
realized as a CFT describing a quantum critical point between
states with N\'eel and valence bond solid (VBS) order.

This paper will propose a critical spin liquid ground state
described by a CFT with a global SO(5) symmetry. We begin with the
SU(4) CFT of Dirac fermions noted above, and examine the
renormalization group (RG) flow of all perturbations which
preserve relativistic invariance and at least a global
SO(5)\,$\subset$\,SU(4) symmetry. The global symmetry of the
underlying antiferromagnet involves only a continuous SU(2) spin
rotation symmetry and various discrete space group symmetries, and
all such SO(5) invariant perturbations will generically be
present. Using an expansion defined below, we find that for
physical parameter values the SU(4) fixed point is unstable to
flow towards a SO(5) invariant fixed point.

Our motivation for examining CFTs with SO(5) symmetry comes from
an interesting proposal by Tanaka and Hu \cite{tanaka} (see also
the work of Senthil and Fisher \cite{sf}). They examined the
quantum fluctuations of the 3-component N\'eel order parameter,
$\boldsymbol{N}$, and the complex VBS order parameter $\Xi$, and
suggested that they be combined into a single five-component real
vector $\Sigma_a$, with $a=1 \ldots 5$ which transforms as the
fundamental of an enlarged SO(5) group. In a spin liquid with such
an SO(5) symmetry, the anomalous dimensions of $\boldsymbol{N}$
and $\Xi$ would be equal, and Sandvik's numerical results
\cite{sandvik} on the quantum critical point between the N\'eel
and VBS states are consistent with such an equality.

We begin by reviewing the SU(4)-invariant CFT of Dirac fermions,
largely following the notation of Ref.~\onlinecite{motherasl}. The
CFT is described by the Euclidean spacetime action $\mathcal{S}_0
= \int d^2 r d \tau \mathcal{L}_0$, where \beqn \mathcal{L}_0 =
\overline{\Psi}^\alpha \gamma^\mu (\partial_\mu + i a_\mu)
\Psi_\alpha \label{s0} \eeqn where $\mu = \tau, x, y$ is a
spacetime index, $\gamma^\mu$ are the Dirac matrices,
$\overline{\Psi} = \Psi^\dagger\gamma_0$ and $a_\mu$ is an
emergent U(1) gauge field. As shown in earlier work
\cite{rantwen,motherasl}, this action defines an SU($N$) invariant
CFT in an expansion in $1/N$. The combined N\'eel-VBS operator,
$\Sigma_a$, can be written in terms of the $\Psi_\alpha$ by \beqn
\Sigma_a = \overline{\Psi} \Gamma_a \Psi \eeqn where $\Gamma_a$
are five $4\times4$ matrices from the SU(4) algebra. This algebra
can be realized using the tensor product of two independent sets
of Pauli matrices, $\boldsymbol{\mu}$ and $\boldsymbol{\sigma}$,
and Hermele {\em et al.\/} showed that $\Gamma_a = (\mu^z
\sigma^x, \mu^z \sigma^y, \mu^z \sigma^z, \mu^x, \mu^y)$. A
curious, and key, property of the $\Gamma_a$ is that they
anti-commute, $\{ \Gamma_a, \Gamma_b \} = 2 \delta_{ab}$, and so
they are Dirac matrices of five spacetime dimensions. The 10
generators of the SO(5) group, under which $\Sigma_a$ transforms
as a SO(5) fundamental, are obtained from the commutators of the
$\Gamma_a$: \beqn \Gamma_{ab} = \frac{1}{2i}[\Gamma_a, \Gamma_b].
\eeqn The $\Gamma_a$ and $\Gamma_{ab}$ are the complete set of
SU(4) generators.

It will be important for our analysis to be able to generalize
these order parameters, and the associated algebraic structure,
from SU(4) to general SU($N$), so as to allow a systematic $1/N$
expansion. A similar strategy was used in the context of chiral
symmetry breaking of three dimensional QED \cite{herbut}. However,
the above embedding of SO(5) into SU(4) relies on the spinor
representations of SO(5), and this does not have a suitable
generalization. However, we note that there is an antisymmetric
matrix $\mathcal{J} = i \sigma^y \mu^x$, with $\mathcal{J}^2 =
-1$, under which \beqn \mathcal{J} \Gamma_{ab} \mathcal{J} =
\Gamma_{ab}^T \label{j} \eeqn for all $ab$. Eq.~(\ref{j}) is the
defining relation for generators of the Sp(4) subgroup of SU(4),
and we have just established the well-known congruence Sp(4)
$\cong$ SO(5). The embedding of Sp($N$) into SU($N$) generalizes
easily to all even $N$, with an $N \times N$ antisymmetric
$\mathcal{J}$ matrix obeying $\mathcal{J}^2 = -1$. We will
therefore study here the SU($N$) invariant CFT in Eq.~(\ref{s0})
with $\alpha=1 \ldots N$, while allowing perturbations which are
invariant under Sp($N$).

A linear stability analysis of this SU($N$) CFT has been carried
out earlier \cite{motherasl} for a limited set of perturbations.
For sufficiently large $N$, all perturbations are believed to be
irrelevant. However, the anomalous dimensions arising at order
$1/N$ can be quite large, and we shall show find below a
perturbation which becomes relevant when its scaling dimension is
evaluated at $N=4$. We are also interested in finding a systematic
approach to determining the fate of such a relevant perturbation,
beyond a linear stability analysis. To this end, we will allow the
tree-level scaling dimensions to vary as a function of spatial
dimensionality, $d$, as is common in other critical phenomena
contexts. With Dirac fermions, there is the subtle issue of
dimensional continuation of the Dirac matrices, $\gamma^\mu$; as
is commonly done \cite{ssl}, we will deal with this by applying
the Dirac algebra and phase space factors as in $d=2$. Our
stablility analysis of spin liquids and their perturbations is
formally justified by taking $(d-1) \propto 1/N$, and then
expanding in $1/N$.

It is also interesting to consider application of this method to
antiferromagnets in $d=1$. Although we will not describe the
computation here, it is necessary to adapt our results to Dirac
matrices in $d=1$. From such a computation, we reproduced the
results of Affleck \cite{affleck} on the spectrum of scaling
dimensions of operators with SU($N$) and Sp($N$) symmetry at the
fixed points described by WNZW models.

We now present our RG results for perturbations of the CFT in
Eq.~(\ref{s0}). We begin by considering perturbations which are
invariant under SU($N$). To the order we are working, there are
only two independent perturbations, which we write as \beqn
\mathcal{L}_1 =  \frac{\lambda_1}{N} (\overline{\Psi}^\alpha
\Psi_\alpha ) (\overline{\Psi}^\beta \Psi_\beta ) +
\frac{\lambda_2}{N} (\overline{\Psi}^\alpha \gamma^\mu \Psi_\alpha
) (\overline{\Psi}^\beta \gamma_\mu \Psi_\beta ) \label{s1} \eeqn
where the circular brackets indicate a trace over indices in Dirac
space. Other possible terms, such as $(\overline{\Psi}^\alpha
\Psi_\beta ) (\overline{\Psi}^\beta \Psi_\alpha )$ and
$(\overline{\Psi}^\alpha \gamma^\mu \Psi_\beta )
(\overline{\Psi}^\beta \gamma_\mu \Psi_\alpha )$, can be shown to
be linearly related to the terms in Eq.~(\ref{s1}).

\begin{figure}
\includegraphics[width=2.8in]{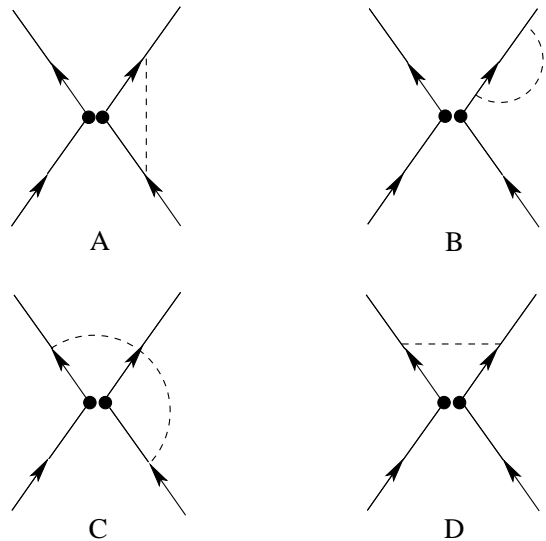}
\caption{Feynman diagrams contribute to the linear orders in both
Eqs. \ref{rgsun} and \ref{rgspn}. The dashed lines are dressed
photon propagators, and the full circles denote the trace in Dirac
space.} \label{fig1}
\end{figure}

\begin{figure}
\includegraphics[width=3.0in]{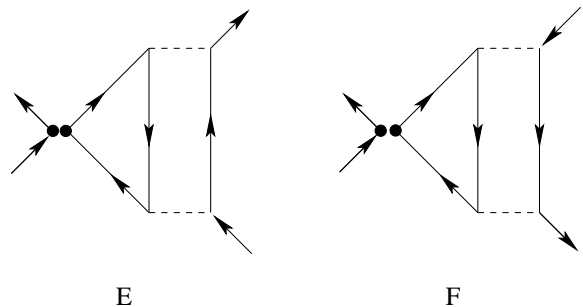}
\caption{Feynman diagrams which only contribute to the linear
orders in Eqs. \ref{rgsun}, but not in Eqs. \ref{rgspn}.}
\label{fig2}
\end{figure}
\begin{figure}
\includegraphics[width=3.0in]{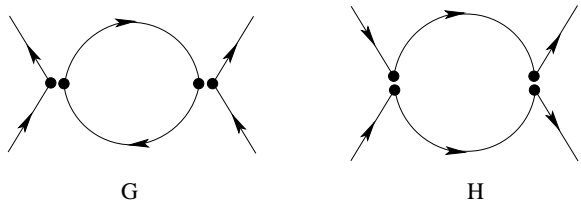}
\caption{Feynman diagrams which contribute to the quadratic order
of the RG equations (\ref{rgsun}) and (\ref{rgspn}). Notice that
since we only calculate to the order of unity in the quadratic
terms, diagram G only contributes to equation (\ref{rgsun}) but
not (\ref{rgspn}), and diagram H only contributes to equation
(\ref{rgspn}) but not (\ref{rgsun}).} \label{fig3}
\end{figure}

From the diagrams shown in Fig.~\ref{fig1}, Fig.~\ref{fig2} and
Fig.~\ref{fig3}, we obtained the following RG equations for a
rescaling by a factor $e^\ell$: \beqn \frac{d\lambda_1}{d \ell}
&=& \left( 1 - d - \frac{256}{3 N \pi^2} \right)\lambda_1 +
\frac{64}{N \pi^2} \lambda_2 -  \frac{2}{\pi^2} \lambda_1^2,
\cr\cr\cr \frac{d\lambda_2}{d \ell} &=& \left( 1-d
\right)\lambda_2 + \frac{64}{3 N \pi^2}  \lambda_1 +
\frac{2}{3\pi^2} \lambda_2^2. \label{rgsun}\eeqn

In the terms linear in the $\lambda$ on the right-hand-side, we
have computed co-efficients to order $1/N$, and the $1/N$
corrections come from the dressed photon propagator \cite{lee1999}
\beqn G_{\mu\nu}(p) = \frac{16}{Np}(\delta_{\mu\nu} - \frac{p_\mu
p_\nu}{p^2}).\eeqn  For the terms quadratic in $\lambda$ to be of
the same order as the linear terms, we need only compute the
co-efficients to order unity, as is the case above.

The RG equations in Eq.~(\ref{rgsun}) have several fixed points,
but we begin by considering the fixed point at
$\lambda_1=\lambda_2=0$. The eigenvalues at this fixed point are
$1-d - (128 \pm 64 \sqrt{7})/(3 N \pi^2)$. At the physical values
of $d=2$ and $N=4$, these eigenvalues evaluate to $-0.651$ and
$-3.510$. So both are negative and the $\lambda_1=\lambda_2=0$
fixed point is stable. None of the other fixed points of
Eq.~(\ref{rgsun}) were found to be stable at these values of $d$
and $N$. By examining the $N$ dependence of the eigenvalues at
$\lambda_1=\lambda_2=0$ we conclude that the SU($N$) CFT defined
by Eq.~(\ref{s0}) is stable to SU($N$)-invariant perturbations for
$N > 1.40/(d-1)$.

Next we consider the additional perturbations of $\mathcal{L}_0$
when the global symmetry is reduced from SU($N$) to Sp($N$). A
simple analysis shows there is only one allowed term \beqn
\mathcal{L}_2 =  \frac{\lambda_3}{N} \mathcal{J}_{\alpha\gamma}
\mathcal{J}^{\beta \delta} (\overline{\Psi}^\alpha  \Psi_\beta )
(\overline{\Psi}^\gamma \Psi_\delta ). \label{s2} \eeqn A second
possible term $\mathcal{J}_{\alpha\gamma} \mathcal{J}^{\beta
\delta} (\overline{\Psi}^\alpha  \gamma^\mu \Psi_\beta )
(\overline{\Psi}^\gamma \gamma_\mu \Psi_\delta )$ reduces to the
above term after application of Fierz identities.

From the diagrams in Fig. \ref{fig1} and diagram H in Fig.
\ref{fig3}, the RG equations for $\mathcal{L}_2$ reads (Notice
that diagrams in Fig. \ref{fig2} and diagram G in Fig. \ref{fig3}
do not contribute to the leading order of 1/N expansion) \beqn
\frac{d\lambda_3}{d\ell} = \left( 1-d + \frac{64}{N \pi^2}
\right)\lambda_3 - \frac{1}{3 \pi^2} \lambda_3^2. \label{rgspn}
\eeqn This has fixed points at $\lambda_3=0$ and $\lambda_3 =
\lambda_3^\ast = 3 \pi^2 (1-d + 64/(N \pi^2))$. At $d=2$ and $N=4$
we now find a result which is very different from the SU($N$)
perturbations above. The $\lambda_3=0$ fixed point is unstable
with RG eigenvalue $0.621$, while the fixed point at $\lambda_3 =
\lambda_3^\ast >0$ is stable; for general $N$, we find that the
stablity of the $\lambda_3 = \lambda_3^\ast$ fixed point holds for
$N<6.48/(d-1)$. So for $1.40/(d-1) < N < 6.48/(d-1)$, the theory
$\mathcal{L}_0 + \mathcal{L}_1 + \mathcal{L}_2$ flows to a fixed
point with $\lambda_1=\lambda_2=0$ and $\lambda_3 =
\lambda_3^\ast$ which describes our advertised Sp($N$)-invariant
critical spin liquid.

\begin{figure}
\includegraphics[width=3.0in]{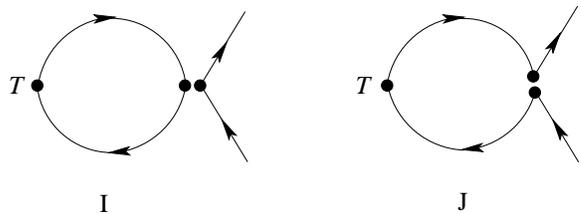}
\caption{Feynman diagrams which contribute to the difference of
scaling dimensions of fermion bilinears
$\overline{\Psi}\Gamma_{a}\Psi$ and
$\overline{\Psi}\Gamma_{ab}\Psi$.} \label{fig4}
\end{figure}

The scaling dimensions of all $16$ fermion bilinears
$\overline{\Psi}T_a\Psi$ ($T_a$ are SU($4$) generators with $a =
1$, $\cdots$ $15$) equal at the fixed point with $\lambda_i = 0$
which respects the SU($4$) symmetry. At the order of $1/N$, the
scaling dimensions read: \beqn \Delta(\overline{\Psi} T_a\Psi) = 2
- \frac{64}{3N\pi^2}, \cr\cr \Delta(\overline{\Psi}\Psi) = 2 +
\frac{128}{3N\pi^2}.\eeqn with $N = 4$, the difference between the
two scaling dimensions above is from the diagrams similar to the
ones in Fig. \ref{fig3} \cite{motherasl} with two photon
propagators and a trace in the fermion flavor space, which only
contributes to fermion bilinear $\overline{\Psi}\Psi$. At the
Sp($4$) symmetric fixed point, the scaling dimensions of fermion
bilinears are classified as the representation of
Sp(4)$\simeq$SO(5) group: $\overline{\Psi}\Psi$,
$\overline{\Psi}\Gamma_a\Psi$ and $\overline{\Psi}\Gamma_{ab}\Psi$
form scalar, vector and adjoint representations of SO(5) group
respectively, and the scaling dimensions of fermion bilinears
within the same representation are equal to each other.

For larger $N$, the scaling dimensions of the fermion bilinears at
the Sp($N$) fixed point deviate from their value at the SU($N$)
fixed point at the order of $1/N^2$, and requires a lot more
calculations. But their differences at $1/N^2$ order can be
calculated readily from diagrams in Fig. \ref{fig4}: \beqn
\Delta(\overline{\Psi}\Gamma_{a}\Psi) -
\Delta(\overline{\Psi}\Gamma_{ab}\Psi) =
\frac{6\lambda^\ast_3}{\pi^2N}. \label{scaling}\eeqn Here
$\Gamma_a$ and $\Gamma_{ab}$ together form a fundamental
representation of SU($N$) algebra, and $\Gamma_{ab}$ form the
spinor representation of Sp($N$) subalgebra.

To fully analyze the physical implications of this fixed point, we
have to examine the fate of all perturbations which further reduce
the global symmetry from Sp(4) down to those required by the SU(2)
spin rotation symmetry and the square lattice space group. There
are a large number of such additional perturbations, and analyzing
them all would require an analysis of daunting complexity. We also
need a procedure for generalizing such perturbations to general
Sp($N$) operators to enable a $1/N$ expansion, and there is no
unique and natural choice like the one we have used so far; the
results will depend upon the particular choices made for the
invariant subgroups of Sp($N$). We will therefore not present such
an analysis here. Additional perturbations which break Lorentz
invariance are also possible; there were examined by Hermele {\em
et al.} \cite{motherasl}, and found to be irrelevant.

Should no relevant perturbations emerge at the Sp(4) fixed point,
it would describe a stable critical spin liquid phase. Otherwise
it would be a (multi-) critical point between ordered phases, with
the dimensionality of the phase diagram determined by the number
of relevant operators. An intriguing possibility is that there is
only one relevant perturbation, which drives the system to a
N\'eel or a VBS state on opposite sides of the Sp(4)-invariant
critical point.

Such a Sp(4) $\cong$ SO(5) fixed point separating N\'eel and VBS
states was suggested by Tanaka and Hu \cite{tanaka}. They further
proposed a SO(5) non-linear sigma model, with a Wess-Zumino term
which could realize that a critical state. However, our Sp(4)
critical point also has a U(1) gauge field, and an associated
conserved topological current, and there is no analog of this
conserved current in the Tanaka-Hu sigma model. So it is likely
that our Sp(4) critical spin liquid is distinct from their
proposal \cite{senthil2}.

A large number of possible spin liquid ground states have been
proposed for the square lattice antiferromagnet. All previous
proposals have been associated with a mean-field saddle point of a
theory of electrically neutral spinons which are either fermions
or bosons. This paper has proposed a novel type of a spin liquid,
which does not have a direct mean-field realization, but is
induced by the gauge fluctuations about a mean-field saddle point.
The only numerical evidence so far of a spin liquid state on the
square lattice for SU(2) antiferromagnets is in the studies of the
transition point between N\'eel and VBS states
\cite{sandvik,rkk,shailesh}. Our SO(5) spin liquid is a candidate
for this state, as it can explain the possible equality of the
scaling dimensions of the N\'eel and VBS operators. A further
testable property of our spin liquid is that the 10 observable
operators \cite{motherasl} defined by $\overline{\Psi} \Gamma_{ab}
\Psi$ all have equal scaling dimensions, which are distinct from
those of the N\'eel and VBS orders.

We thank T.~Senthil for valuable discussions. This work was
supported by NSF Grant No.\ DMR-0537077.

\end{document}